\documentclass[aps,prl,twocolumn,superscriptaddress]{revtex4}

\usepackage{graphicx}

\usepackage{amssymb}
\usepackage{amsmath}
\usepackage{bm}
\usepackage{color}

\newcommand{\lla}{\left\langle}
\newcommand{\rra}{\right\rangle}

\DeclareFontFamily{U}{mathc}{}
\DeclareFontShape{U}{mathc}{m}{it}%
{<->s*[1.03] mathc10}{}

\DeclareMathAlphabet{\mathscr}{U}{mathc}{m}{it}

\graphicspath{{figures/}}

\begin{document}

\title{Enhanced rotational motion of spherical squirmer in polymer solutions}

\author{Kai Qi}
\email{k.qi@fz-juelich.de}
\affiliation{Theoretical Soft Matter and Biophysics, Institute of Complex
Systems and Institute for Advanced Simulation, Forschungszentrum J\"ulich,
52425 J\"ulich, Germany}
\author{Elmar Westphal}
\email{e.westphal@fz-juelich.de}
\affiliation{Peter Gr\"unberg Institute and J\"ulich Centre for Neutron Science,
Forschungszentrum J\"ulich, D-52425 J\"ulich, Germany}
\author{Gerhard Gompper}
\email{g.gompper@fz-juelich.de}
\affiliation{Theoretical Soft Matter and Biophysics, Institute of Complex
Systems and Institute for Advanced Simulation, Forschungszentrum J\"ulich,
52425 J\"ulich, Germany}
\author{Roland G. Winkler}
\email{r.winkler@fz-juelich.de}
\affiliation{Theoretical Soft Matter and Biophysics, Institute of Complex
Systems and Institute for Advanced Simulation, Forschungszentrum J\"ulich,
52425 J\"ulich, Germany}

\begin{abstract}
The rotational diffusive motion of a self-propelled, attractive spherical colloid immersed in a solution of self-avoiding polymers is studied by mesoscale hydrodynamic simulations. A drastic enhancement of the rotational diffusion by more than an order of magnitude in presence of activity is obtained. The amplification is a consequence of two effects, a decrease of the amount of adsorbed polymers by active motion, and an asymmetric encounter with polymers on the squirmer surface, which yields an additional torque and random noise for the rotational motion. Our simulations suggest a way to control the rotational dynamics of squirmer-type microswimmers by the degree of polymer adsorption and system heterogeneity.

%The rotational diffusion coefficient exhibits a nonmonotonic dependence on the polymer concentration, it increases %first with increasing concentration and decrease again above a critical value. The polymer conformations are affected %by activity, with polymer swelling by activity at smaller concentrations. This reflects the viscoelastic nature of %the polymer solution and intimate coupling of polymer deformations and activity.

%An attractive polymer-colloid interaction leads to enhancement of rotational diffusion at polymer concentrations %below the critical value and a reduction above that value. Yet, the rotational motion is accelerated compared to the %passive colloid under the same conditions, because the passive rotational diffusion coefficient decreases %substantially with increasing polymer concentration and attraction strength.

\end{abstract}

\maketitle

%%%$\mathscr{abcdefghijklmnopqrstuvwxyz}$

Self-propelled microorganisms are habitually exposed to complex fluid environments \cite{Elgeti2015} consisting of solutions with a broad range of dispersed macromolecules and colloidal particles \cite{McGuckin2011}. However, our current understanding of microorganism locomotion mainly rests upon Newtonian fluids whose properties are governed by viscous stresses, whereas complex fluids are viscoelastic, i.e., they are non-Newtonian, which implies additional elastic effects \cite{laug:07.1,Patteson2015,Qiu2014}. Potential medical and industrial applications have triggered numerous investigations of the motility of microorganisms in complex fluids, like bacteria swimming and swarming  in a biofilm \cite{Houry2012} and moving sperm in the reproductive tract \cite{Fauci2006}.
%Viscoelasticity not only stem from the peculiarities of viscoelastic fluids, such as shear thinning and thickening \cite{Qiu2014}, but also depend on the specialties of the swimmers like body geometry \cite{Spagnolie2013}.
Intuitively, the existence of high-molecular weight macromolecules can be expected to slow down the translational motion of swimmers because of the (substantially) enhanced viscosity \cite{Shen2011,Zhu2012,Qin2015,datt:17}. However, increased swimming speeds have been reported  \cite{Berg1979,Teran2010,Espinosa2013,Patteson2015,Martinez2014,Jung2010,Leshansky2009,Gomez2017,Magariyama2002,Zottl2019}, where enhancement is attribute to mechanical responses caused by fluid viscoelasticity \cite{Teran2010,Espinosa2013,Riley2014,Patteson2015,datt:17}, local shear thinning \cite{Martinez2014,Spagnolie2013}, and polymer depletion\cite{Man2015,Zottl2019}.
%In the meanwhile, enhancements and hindrances can also be achieved simultaneously in some systems by tuning fluids and swimmers properties properly \cite{Liu2011,Spagnolie2013,Thomases2014,Datt2015,De2015,Schneider1974}.
In addition, viscoelasticity affects other microswimmer properties, such as their rotational motion.  Recent experimental studies of self-propelled Janus colloids in a viscoelastic fluid yield a drastically enhanced rotational diffusion by up to two orders of magnitude \cite{Gomez2016}. A further increase of activity can even result in persistent rotational motion \cite{Narinder2018}.
%The enhancement has been attributed to memory-related friction forces which oblique to instantaneous orientation of a microswimmer.
%Yet, the microscopic mechanism of the enhanced rotational diffusion can not be unambiguously revealed experiments. But %computer simulations can tackle this difficult challenge.

In order to shed light onto microscopic mechanisms that lead to an enhanced rotational motion, we perform mesoscale hydrodynamic simulations of a spherical squirmer embedded in a  fluid employing the multiparticle collision dynamics (MPC) approach \cite{Kapral2008,Gompper2009}. Viscoelasticity is captured by taking linear polymers  explicitly into account, which we consider to adsorb on the colloid surface.
By variation of the squirmer activity, our simulations yield a drastic enhancement of its rotational diffusion by more than an order of magnitude, in particular in a dilute solution of self-avoiding polymers. This increase is induced by an inhomogeneous distribution of (partially) adsorbed polymers on the squirmer surface moving in the squirmer-induced flow field. Through the adjustment of the squirmer-polymer-interaction strength and system heterogeneity, our results demonstrate the feasibility of controlling the rotational motion of squirmer-type microswimmers in a complex environment.

\begin{figure}
\centerline{\includegraphics[width=0.8\columnwidth]{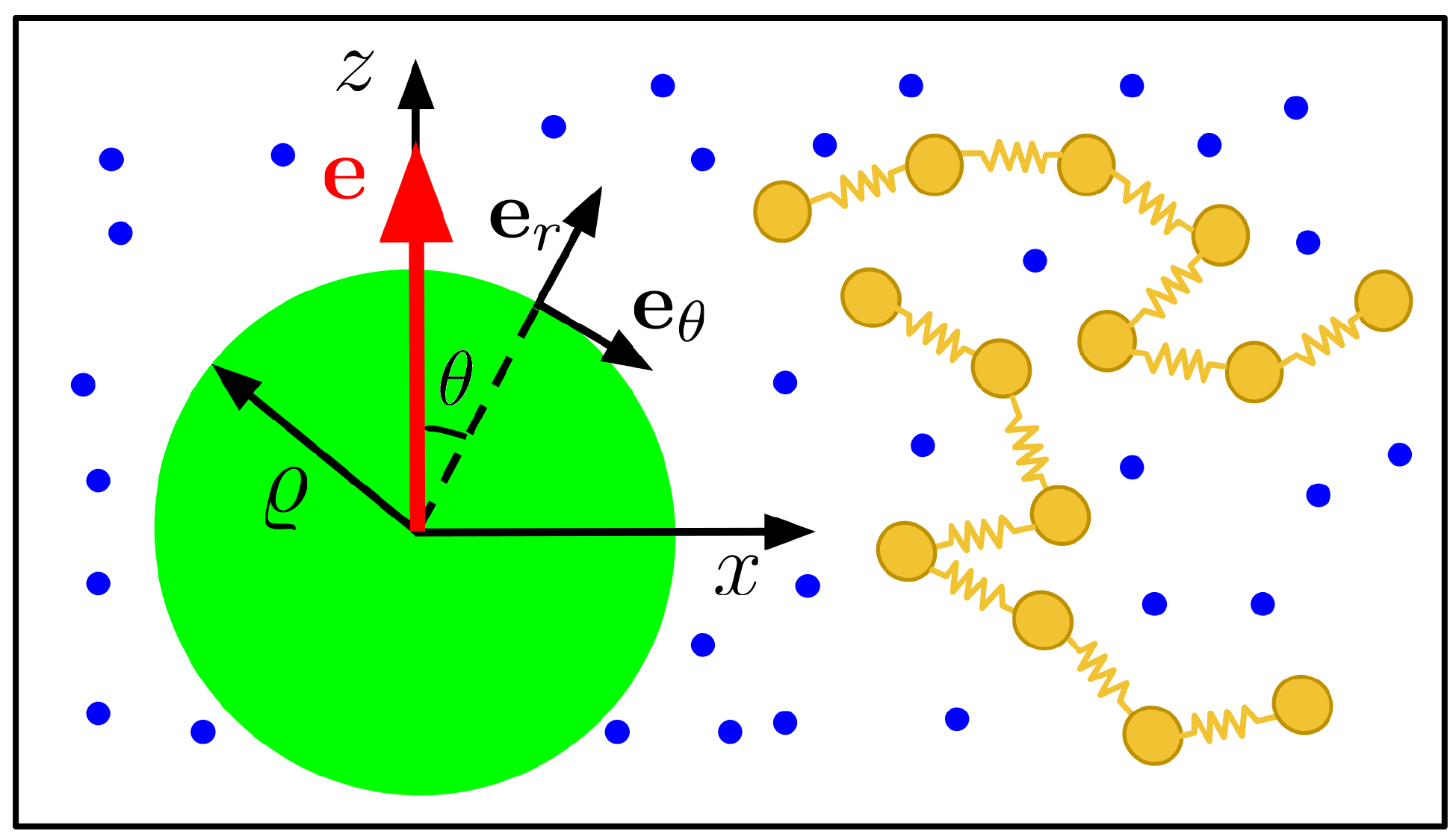}}
\caption{\label{Fig1}%
Sketch for a spherical squirmer of radius $\varrho$ with propulsion direction $\bm e$, radial unit vector $\bm e_r$, and tangential unit vector $\bm e_{\theta}$  immersed in a viscoelastic fluid of  MPC fluid particles (blue dots) and self-avoiding polymers (yellow bead-spring chains).}
\end{figure}

We consider a squirmer, which is modeled as a neutral buoyant hard sphere of radius $\varrho$ with the prescribed tangential surface slip velocity \cite{Ishikawa2006,Theers2016}
\begin{equation} \label{eq:surf_vel}
    {\bm u}_{sq}= \frac{3}{2} U_0 \sin(\theta) \left(1+\beta \cos(\theta) \right) {\bm e}_{\theta} ,
\end{equation}
where $U_0$ is the swimming velocity, $\theta$ is the polar angle with respect to the squirmer's orientation and swimming direction ${\bm e}$, ${\bm e}_{\theta}$ is the local tangent vector, and $\beta$ characterizes the active stress ($\beta <0$ pusher, $\beta=0$ neutral squirmer, $\beta>0$ puller), see Fig.~\ref{Fig1}. The actual flow field of a swimmer is determined by its particular propulsion mechanism, which for a diffusiophoretic particle depends on the details of the spatial distribution and the type of chemical reactions across its surface \cite{pope:18}.  Here, we focus on neutral squirmer, i.e., $\beta=0$.  Results for pushers and pullers are presented in the Supplemental Material (SM) \cite{supp}. 

A linear polymer is composed of $N_m$ touching beads, connected by strong harmonic bonds of finite rest length, $l_0$, [Fig.~\ref{Fig1}] (cf. SM for details).
Polymer excluded-volume interactions are taken into account by a truncated and repulsive Lennard-Jones potential. Polymer adsorption onto the squirmer surface is triggered by the radial,
attractive separation-shifted Lennard-Jones-type potential
\begin{equation}
\frac{U_a(r_a)}{k_BT}=4 \epsilon_a \left[ \left( \frac{a}{r_a+1.377 a} \right)^{8}-\left(\frac{a}{r_a+1.377 a}\right)^4 \right],
\end{equation}
for monomer-squirmer distances $r_a < 3 a$  and zero otherwise, where $r_a$ is the monomer distance with respect to the colloid surface, $\epsilon_a$ the attraction strength, and $a$ the length of a MPC collision cell (cf. SM). At the colloid surface, the monomers experience an effective
slip velocity due to the tangential squirming velocity of the fluid.
The latter can be understood as an effective description of the transport mechanism of
ciliated or phoretic microswimmers. A well-known example is the transport of mucus
(a viscous polymer gel) by ciliated surfaces in the airways. Similarly, we expect
polymer adsorption on colloids to be generic, as for polyacrylamide
polymers absorbing on both halves of the silica-carbon Janus particles employed in  Ref.~\cite{Gomez2016}.
By the coarse-grained nature of our polymer model, every monomer bead corresponds to
several molecular segments of a real polymer,
with a correspondingly enhanced attraction strength. However, the ratio between colloid diameter
$2\varrho$ and bead size $\sigma \approx l_0$ (typically $2\varrho/l_0=12$) is small compared to that of a real
colloid and polymer. Nevertheless, the qualitative behavior does not depend significantly on the monomer size,
as simulations of systems of phantom polymers with point-like beads reveal a qualitative
similar behavior as of self-avoiding polymers with Lennard-Jones beads.

%\begin{equation}
%\frac{U_a(r_a)}{k_BT}=4 \epsilon_a \left[ \left( \frac{a}{r_a+1.377 a} \right)^{8}-\left(\frac{a}{r_a+1.377 a}\right)^4 %\right]
%\end{equation}
%for monomer-squirmer distances $r_a < 3 a$  and zero otherwise, where $r_a$ is the monomer distance with respect to the colloid surface, $\epsilon_a$ is the attractive strength in units of $k_BT$, and $a$ is the length of a MPC collision cell (see below). The shift $1.377 a$ ensures that the strongest attraction is on the surface.

The squirmer and polymer dynamics is treated by molecular dynamics simulations (MD), describing the rotational motion of a squirmer by quaternions \cite{Theers2016}. For the MPC fluid \cite{Kapral2008,Gompper2009},
the stochastic-rotation-dynamics variant with angular momentum conservation (MPC-SRD+a) is applied \cite{thee:16}. Details of the implementation and the applied parameters are presented in the SM.

We performed between 25 and 50 independent simulation runs of $5 \times 10^6$ MPC steps ($10^8$ MD steps) for every displayed parameter set. Averages, denoted by $\langle \cdots \rangle$, are taken over the various realization and well separated configurations of individual runs (time average).

%A squirmer of radius $R=6a$ is placed in a  cubic box of side length $L=60a$ with periodic boundary conditions. The fluid and squirmer mass density is $\rho=10m/a^3$; $m$ is the mass of a MPC particle. We utilize the MPC time step $h=0.05a\sqrt{m/k_BT}$ and rotation angle $\alpha=130^{\circ}$. For polymers, we set $\kappa=500 k_BT/a^2$, $\sigma_m=0.8a$, $\epsilon_a=15$, and $l_0=a$. The MD time step is $h/20$. The three squirmer velocities   $U_0/\sqrt{k_BT/m} =0, 1/30$, and $1/15$ are considered.

\begin{figure}
\centerline{\includegraphics[width=0.9\columnwidth]{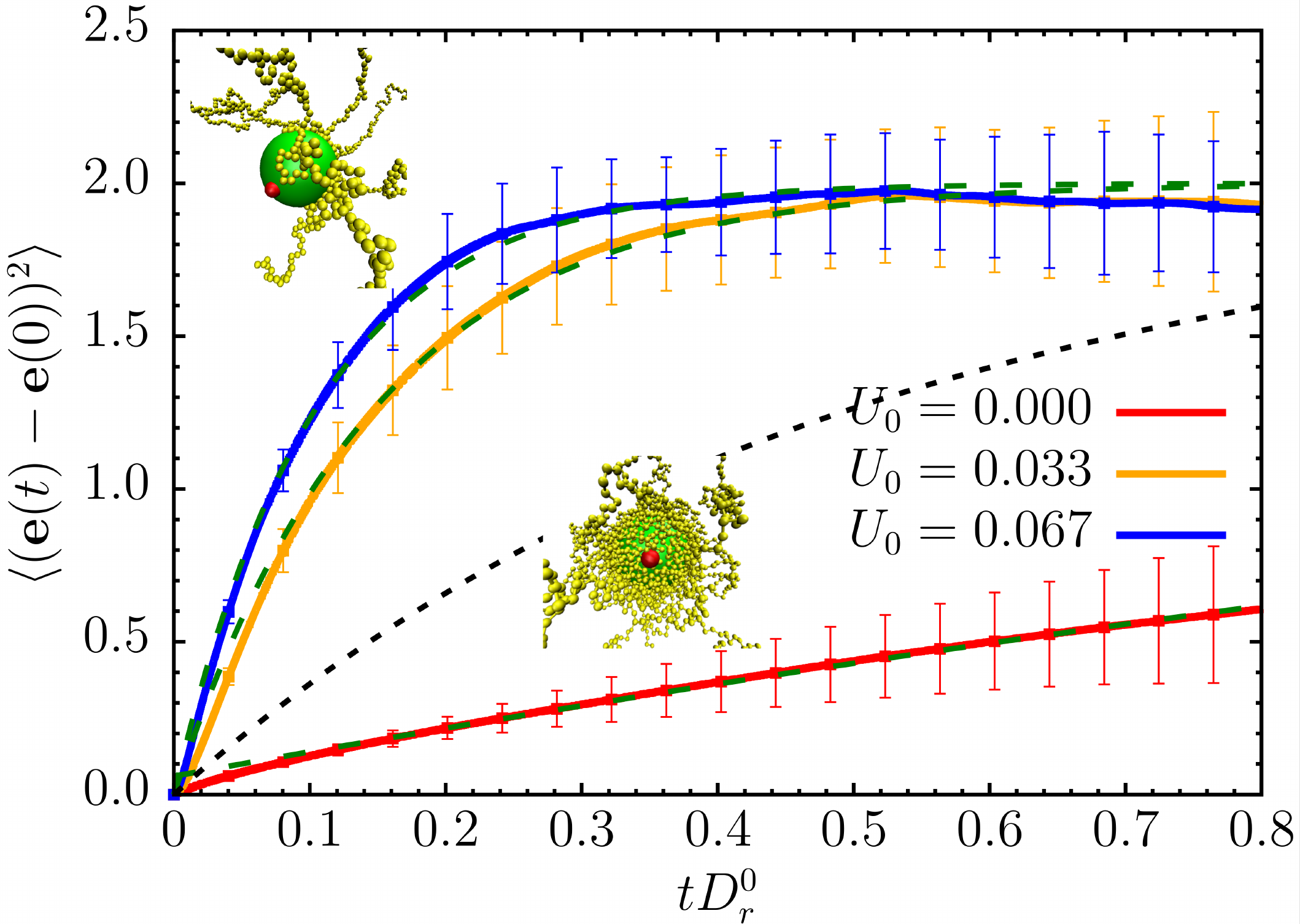}}
\caption{\label{Fig2}%
Rotational mean square displacement (RMSD) of a neutral squirmer, $\beta=0$, as a function of time  for various self-propulsion velocities $U_0$. Green dashed curves are exponential fits according to Eq.~\eqref{eq:rmsd}, which yields the effective rotational diffusion coefficients $D^{a}_r \times 10^{5}/\sqrt{k_BT/ma^2} =0.47,\  7.8, \ 11 $ for $U_0/\sqrt{k_BT/m} = 0, \ 1/30, \ 1/15$, respectively. The black dashed line is the RMSD of a squirmer in a simple fluid, with the rotational diffusion coefficient $D^{0}_r=2.3\times 10^{-5} \sqrt{k_BT/ma^2}$. The error bars indicate the standard deviation. The snapshots illustrate the polymer conformations and distribution.
%(b) Radial monomer density distributions as a function of the monomer-squirmer distance for various $U_0$.
The polymer length is $N_m=240$, the number of polymers $N_p=24$, and the  monomer packing-fraction ratio $\phi/\phi^*= 0.75$. }
\end{figure}

Figure \ref{Fig2} displays simulation results for the rotational mean square displacement (RMSD) of the  propulsion direction $\bm e$ of a neutral squirmer as a function of time for a system with $N_p=24$ polymers of length $N_m=240$ and the packing fraction $\phi/\phi^*=0.75$. Here,  $\phi^*=N_m/V_p$ is the overlap concentration with the polymer volume $V_p=4 \pi R_g^3/3$ in terms of the polymer radius of gyration $R_g$.
Simulations yield $R_g^{0}/l_0 \approx 11.7$ for $N_m=240$ at infinite dilution. By fitting the expression
\begin{align} \label{eq:rmsd}
\lla \left( \bm e(t) - \bm e(0) \right)^2 \rra = 2 \left( 1-e^{-2D^{a}_r t} \right) ,
\end{align}
the activity-dependent rotational diffusion coefficient $D^{a}_r$ can be deduced.
The rotational diffusion coefficient of the passive colloid, $D_r^p = 4.7\times 10^{-6} \sqrt{k_BT/ma^2}$, in the polymer solution is reduced by a factor of five compared to that of the colloid in the bare MPC fluid $D^{0}_r = 2.3\times 10^{-5} \sqrt{k_BT/ma^2}$, reflecting a strong interaction with the polymers. Evidently,  activity significantly enhances the rotational diffusion, which we characterize by the rotational diffusion enhancement $\gamma^{a}=D^{a}_r/D^{p}_r$. Our simulations yield the values $\gamma^{a}=17$ and $\gamma^{a}=23$ for $U_0/ \sqrt{k_BT/m}=1/30$ and $1/15$, respectively.

\begin{figure}
\centerline{\includegraphics[width= 0.75 \columnwidth]{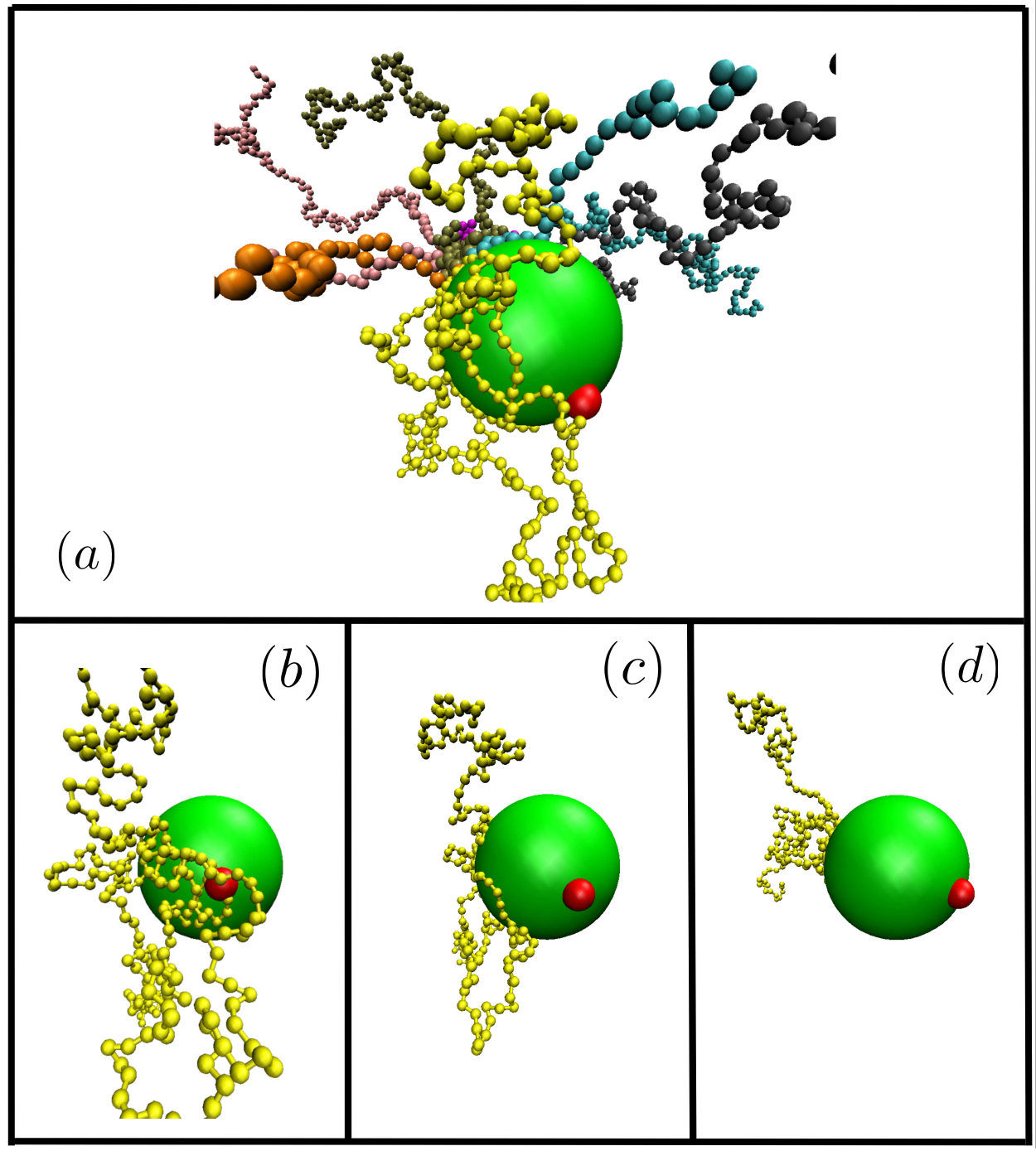}}
\caption{\label{Fig3}%
(a) Snapshot of a neutral squirmer with several adsorbed polymers. The self-propulsion velocity is $U_0/ \sqrt{k_BT/m}=1/15$ and the monomer concentration is $\phi/\phi^*=0.75$. For a better visualization, adsorbed polymers are displayed with different colors.
%(b) Illustration of the squirmer (counter)-rotation by the torque and transport of a particle via the slip velocity %$\bm u_{sq}$.
(b)--(d) Time sequence of the polymer transport and squirmer rotation  by the fluid flow. The red dot indicates the squirmer's propulsion direction.
See  Supplemental Material for an animation.}
\end{figure}

The short-range attraction implies a substantial monomer increase in the vicinity of the colloid, as illustrated in  Fig.~S2, which is the origin of the reduced rotational diffusion of the passive colloid. Activity reduces the monomer concentration, i.e., depletes the polymer next to the squirmer. We attribute this effect to a flow-induced desorption of polymers by the active colloid motion.
Qualitatively, this is similar to experimental \cite{Martinez2014} and simulation results \cite{Zottl2019} for bacteria swimming in a viscoelastic fluid and is assumed to be relevant for their observed enhanced swimming speed. Hence, the increase of the rotational diffusion coefficient can be partially attributed to the decreased amount of adsorbed polymers. %Yet, $D_r^a$ is still smaller than $D_{r}^{0}$ due to the finite amount of adsorbed polymer.

The main rotational enhancement is caused by a second effect, the active polymer transport on the colloid surface.
The directed active motion along $\bm e$ leads to an asymmetric encounter with the dissolved polymers compared to thermal motion---it is enhanced in the propulsion direction (front). At the same time, the imposed slip velocity [Eq.~\eqref{eq:surf_vel}] causes a transport of the polymer from the front of the colloid to its rear.  The polymer transport, together with the required momentum conservation, leads to an enhanced squirmer rotational diffusive motion, as is illustrated in Fig.~\ref{Fig3} for a neutral squirmer. In Fig.~\ref{Fig3}(a), several adsorbed polymers are shown, where the yellow polymer just adsorbs in front. Evidently, the polymer distribution is asymmetric. Figures~\ref{Fig3}(b)-(d) illustrate the dynamics of the yellow polymer, which is gradually transported by flow from the front to the rear part of the squirmer. Momentum conservation implies the opposite motion of the whole squirmer, as is visible by the displacement of the red label.
%This is illustrated in Fig.~\ref{Fig3}(b). The small red point moves with the velocity $\bm u_{sq}$ along the surface %in the counterclockwise direction. Momentum conservation with the opposite force implies the  torque
%\begin{align}
%    {\bm T}_{sq} = - {\bm R} \times (2m {\bm u}_{sq}), \label{squirming_torque}
%\end{align}
%which generates a clockwise rotation of the squirmer.
Since a squirmer more frequently encounters polymers in front, with no preferential azimuthal angular dependence of absorption,
an additional random rotational motion is obtained on top of the rotation by Brownian noise, which enhances the overall rotational diffusion.

\begin{figure}
\centerline{\includegraphics[width= 0.9\columnwidth]{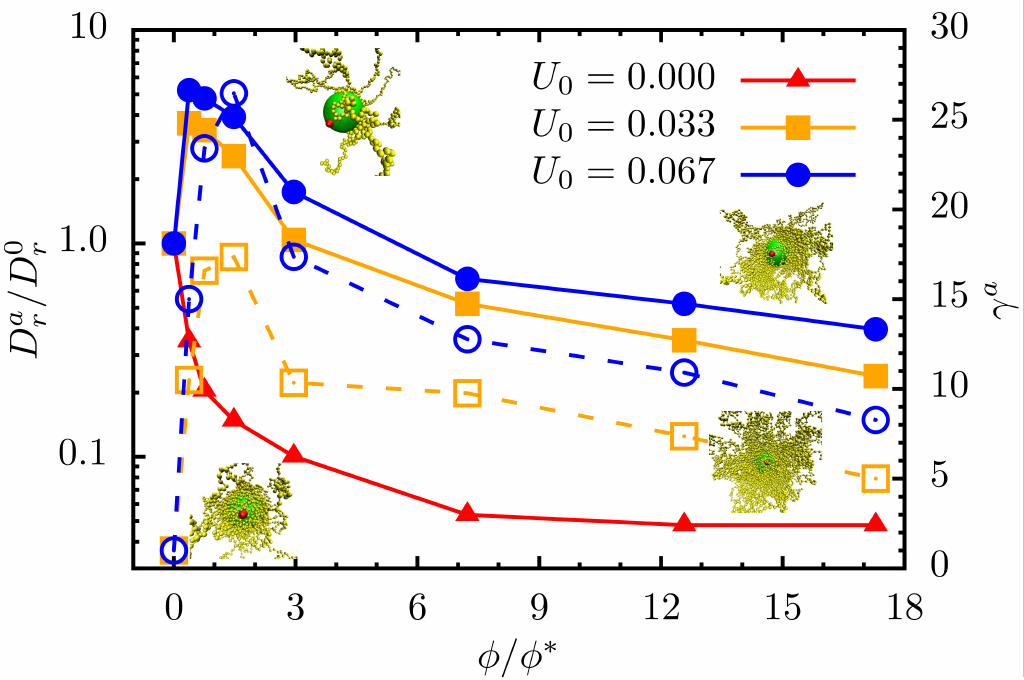}}
\caption{\label{Fig4}%
Normalized rotational diffusion coefficient $D_r^{a}/D_r^0$ (solid lines) and its enhancement $\gamma^a=D_r^a/D_r^p$ (dashed lines) as a function of monomer concentration $\phi/\phi^*$ for the propulsion strengths $U_0/\sqrt{k_BT/m} = 0$ (triangles), $1/30$ (squares), and $1/15$ (bullets); $\beta=0$ and $\epsilon_a=15$. Diffusion coefficients are obtained within 10-25\% accuracy, depending on polymer density. }
\end{figure}

This mechanism is most effective as long as the polymer distribution on the squirmer surface is sufficiently low. At high polymer concentrations, the adsorption rate is larger and more homogeneous with a virtually radially symmetric polymer transport on the surface, which restores a symmetric distribution of extra torques and, hence, implies a drop of the activity-enhanced rotational motion.
Rotational mean square displacements for a neutral squirmer at various swimming velocities are presented in Fig.~\ref{Fig4} as function of the  polymer concentration. (Results for $\beta \ne 0$ are presented and discussed in the SM; Fig.~S7.)  The data for $U_0=0$ indicate the drop of the RMSD by approximately one order of magnitude with increasing polymer concentration. In contrast, for $U_0>0$, the RMSD increases first with increasing concentration and drops for $\phi/\phi^* \gtrsim 0.38$. At small concentrations, $D_r^{a}$ is larger than $D_r^{0}$ of the colloid in the bare MPC fluid. However, above an activity-dependent concentration, $D_r^{a}$ drops below $D_r^0$ ($D_r^a/D_r^0 < 1$ in Fig.~\ref{Fig4}). The diffusion enhancement $\gamma^a=D_r^a/D_r^0$ is significant for all concentrations, despite the fact that $D_r^{a}$ drops below $D_r^{0}$, because  $D_r^p$ decreases stronger with increasing $\phi$ than  $D_r^{a}$. As shown in Fig.~\ref{Fig4}, $\gamma^{a}$ assumes values in the range $1 < \gamma^a \lesssim 25$ for the considered propulsion velocities.  The maximum values are reached at $\phi/\phi^* \approx 1.5$. At $\phi/\phi^* \approx 17$, the rotational enhancement is still significant, with  $\gamma^{a} \approx 5$ and $8$ for $U_0/ \sqrt{k_BT/m}=1/30$ and $1/15$, respectively.

\begin{figure}
\centerline{\includegraphics[width= 0.9\columnwidth]{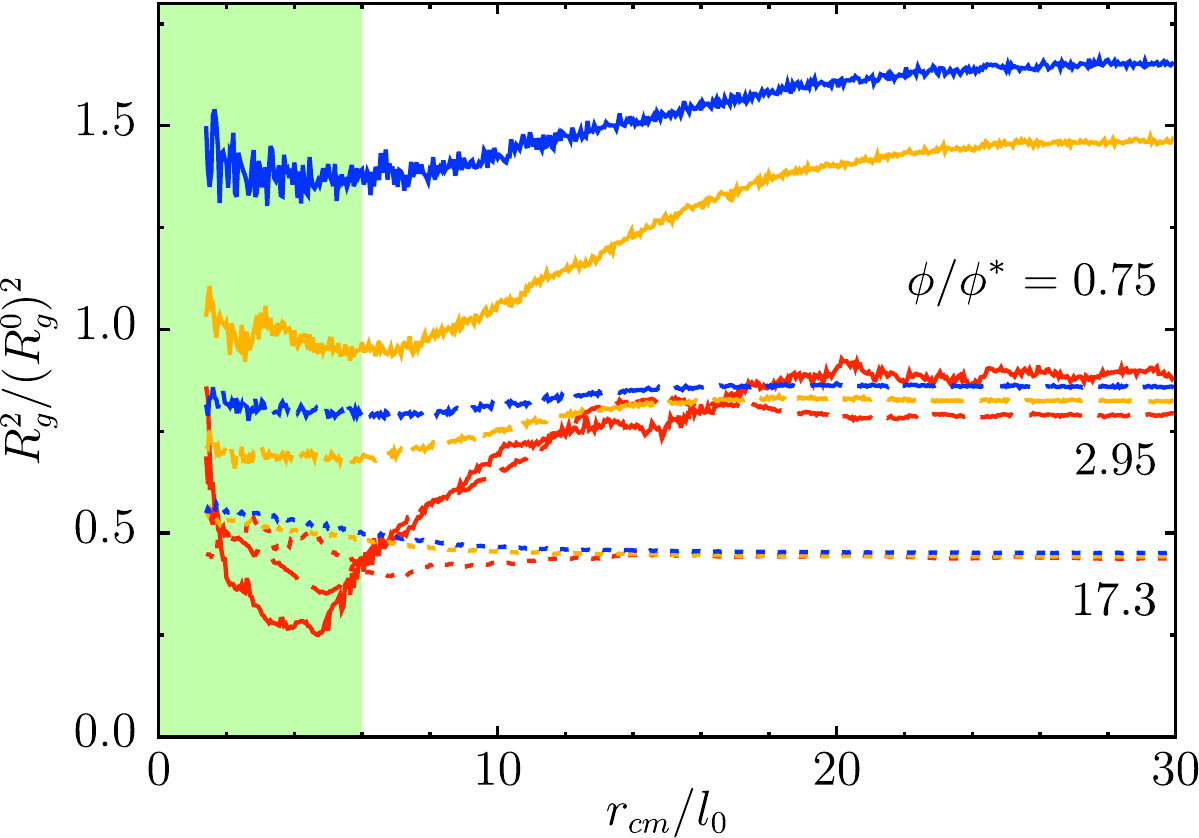}}
\caption{\label{Fig6}%
Polymer radius of gyration $R_g^2$ scaled by the value at infinite dilution, $(R_g^{0})^2$, as a function of the radial polymer center-of-mass position $r_{cm}$ from the squirmer center for $U_0/\sqrt{k_BT/m} =0$ (red), $1/30$ (yellow), and $1/15$ (blue) and the concentration ratios $\phi/\phi^* = 0.75$ (solid), $2.95$ (dashed), and $17.3$ (dotted); $\beta=0$ and $\epsilon_a=15$. The shaded area indicates polymer center-of-mass positions inside of the squirmer.}
\end{figure}

The squirmer locomotion in turn affects the polymer conformations, reflecting the viscoelastic nature of the solution. As shown in Fig.~\ref{Fig6}, the polymer radius of gyration in the presence of a passive colloid decreases with increasing concentration due to screening of polymer excluded-volume interactions ($R_g^2/(R_g^0)^2 <1$) \cite{Doi1986,Huang2010.1}. In fact, for $\phi/\phi^* \gg 1$, the same concentration dependence is obtained, independent of activity (see SM). However, we observe a significant increase in the polymer size at smaller concentrations, $\phi/\phi^* < 3$, which depends on the activity of the squirmer (see also Fig.~S3) \cite{eise:16,wink:17}. Here, due to the limited system size, all polymers are affected. For sufficiently larger system sizes, far from the squirmer, the unperturbed, concentration-dependent equilibrium polymer configuration is assumed (see  SM). The latter is visible for higher concentrations, where the polymer conformations are rather similar for all activities.  In the vicinity of a surface,  the radius-of-gyration of a polymer decreases with decreasing distance from the surface, where surface attraction substantially enhances the effect, whereas activity reduces compactification.  By wrapping around the squirmer, the polymer center-of-mass can be located inside the colloid. Center-of-mass distances $r_{cm}/ \varrho \ll 1$ appear for rather deformed polymers, where the radius of gyration increases again.

Our results indicate a major influence of the polymeric nature of the adsorbed object on the enhancement of the rotational diffusion of squirmers, because simulations of monomers only yield approximately the same squirmer rotational diffusion as obtained for a bare MPC fluid, despite of strong monomer adsorption and active transport (cf. SM for results of polymers with $N_m=60$).

A significant larger enhancement of the rotational diffusion coefficient  has been obtained in the experiments of Ref.~\cite{Gomez2016} for Janus particles in polymer solutions. This could be related to the larger size ratio of the Janus particle and polymer radius of gyration in the experiment compared to the current simulation study. In fact, in experiments the rotational diffusion coefficient of the passive Janus colloid in the polymer solution is by a factor of approximately $30$ smaller than that of the colloid in the bare binary mixture, while in our case, the factor is approximately $5$ at $\phi/\phi^*=0.75$. Hence, a possible depletion of polymer in the vicinity of the colloid by activity would already imply a significantly larger rotational diffusion in experiments than in simulations.

In Ref.~\cite{Narinder2018}, the rotational enhancement is attributed to memory effects of the viscoelastic fluid, where an internal force nonaligned with the actual squirmer-orientation emerges in a coarse-grained continuum ''macroscopic'' description. In simulations, we find concentration-dependent reduced swimming velocities on the order of $40\%$, but the swimming direction is closely aligned with the propulsion direction. Hence, we cannot explain the observed effect by misalignment of propulsion and swimming direction.

In our microscopic simulations, rotational enhancement of neutral squirmers originates from two effects: (i) reduction of the amount of adsorbed polymers by activity compared to that of a passive colloid---implying an increase in the rotational diffusion coefficient---, and (ii) an asymmetric encounter of the squirmer with  polymers at the front  leading to an additional torque by the surface fluid flow, which yields an additional random contribution to the rotational motion. We like to emphasize that the polymer character of the solute is important and that the rotation enhancement depends on polymer length as shown in the SM. In particular, we don't observe an enhancement for a pure monomer solution.

Various other scenarios of colloid-polymer interactions are possible: local attachment with finite life time, temporary local attachment on various patches, local permanent attachment, or even active stress-induced hydrodynamic accumulation ($|\beta| >0$). We present results for the latter two cases in the SM. For a few locally bound beads, which are not transported by the surface flow field, we find a continuous rotational motion opposite to the case of homogeneous attraction, and the squirmer propagates along circular trajectories reminiscent to those observed in Ref.~\cite{Narinder2018}. Sufficiently strong pusher flow fields lead to an enhanced polymer accumulation, which amplifies the rotational diffusive motion (cf. SM for more details).  

The flow field of a phoretic Janus particle is well described by a squirmer-type model employing a phoretic-slip hydrodynamic boundary condition, equivalent to our approach, however, with a somewhat different flow field \cite{pope:18,mich:14}. Hence, the proposed monomer transport by surface slip may also be relevant for Janus particles. The two different materials, however, may lead a preferred adsorption of the polymer to one of them. This could lead to pinning of polymer at their interface line, and consequently to a rotation as emerging by our pinned polymer. Here, further studies taking into account the Janus-type structured colloidal surfaces are desirable, which can be performed by our modelling approach.

Our simulations suggest a possible mechanism to modify and control the rotational motion of an active colloid with slip velocity in polymer solutions. In particular, they emphasize the importance of an inhomogeneous and anisotropic squirmer-polymer interaction. Further studies are necessary to resolve the actual mechanism in experiments, with a rather complex interplay of phoretic flow fields and polymer colloid interactions.

This work has been supported by the DFG priority program SPP 1726 ``Microswimmers – from Single Particle Motion to Collective Behaviour''. The authors gratefully acknowledge the computing time granted through JARA-HPC on the supercomputer JURECA at Forschungszentrum J\"ulich.

%\bibliography{polymer,/Users/winkler/ownCloud/publications_library/bibliography/bibliography}


\end{document}